\documentclass[11pt]{article}
\usepackage{epsfig,amsmath,ctagsplt,subeqnarray,subfigure,tables,float,floatfig,times}
\usepackage[english]{babel}
\usepackage{fancyhdr}

\begin{document}
\pagestyle{empty}

\begin{center}

\bigskip 
\vspace{5cm}

\begin{LARGE} 
\textbf{Associated production 
 \\ of a Z boson and a b-jet in ATLAS } 
\end{LARGE}
   
\vskip 2 cm
   \begin{small}
S. Diglio{\small$^{^{{\,a}{\,b}}}$}, A. Farilla{\small$^{^{\,b}}$}, A. Tonazzo{\small$^{^{{\,a}{\,b}}}$}, 
M. Verducci{\small$^{^{\,c}}$}.\\
\vspace{0.5cm}
{\small{$^{a}$}} {\sl{Dipartimento di Fisica, Universita' Roma Tre and
    INFN }},\\   
{\small{$^{b}$}} {\sl{INFN Sezione di Roma Tre}},\\ 
{\small{$^{c}$}}  {\sl European Organization for Nuclear Research
  (CERN) and CNAF Bologna}.\\  

  \end{small}
 \vskip 3. cm


\end{center}

\begin{center}
 \begin {large} 
 \textbf{Abstract} \end {large} 

 \end{center}
 
\vspace{1.0cm}
{
The current uncertainty on the parametrization of the partonic content of the
proton (PDF's) affects the potential for the discovery of new physics at LHC.
The study of Z boson production in association with a b-jet can considerably
reduce such uncertainty. In addition, this process represents a background
both to the search for the Higgs boson and for SUSY particles.
We present an update, based on the full
simulation data sample produced for the 
Rome Physics Workshop, of a preliminary study \cite{nota} 
in the case where the Z boson decays in $\mu^{+} \mu^{-}$. 

 }

\newpage


\pagestyle{plain}
\pagenumbering{arabic}
\setcounter{page}{1}


\section{Introduction}
The production of electroweak gauge bosons ($W^\pm, Z, \gamma$)
together with jets containing heavy-quarks (c, b) is an important
signal at the hadron colliders.
The simplest process involves a boson and one heavy-quark jet.
In this work, the production of Z boson will be taken into account, 
$gQ\rightarrow ZQ$ (where Q is a b or c quark). The cross sections at the
next-leading-order, see \cite{maltoni1}, will be used to estimate the number of expected events
and finally a comparison with the Tevatron experiment will be done.

\begin{figure}[h!]
 \begin{minipage}[b]{7cm}
   \centering
   \includegraphics[width=7cm]{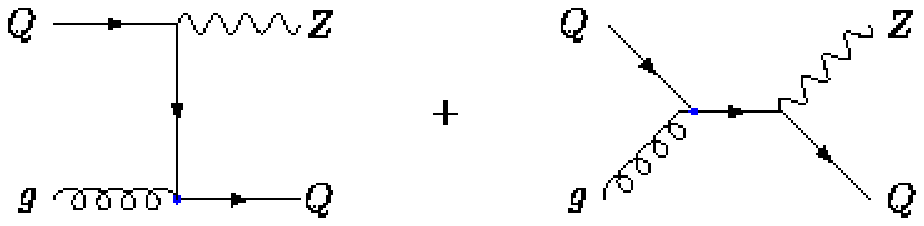}
   \caption{\it Associated production of a Z boson and a single
  hight-$p_T$ heavy quark (Q=c,b).\label{fey}}
 \end{minipage}
 \ \hspace{2mm} \hspace{3mm} \
 \begin{minipage}[b]{7cm}
  \centering
   \includegraphics[width=7cm]{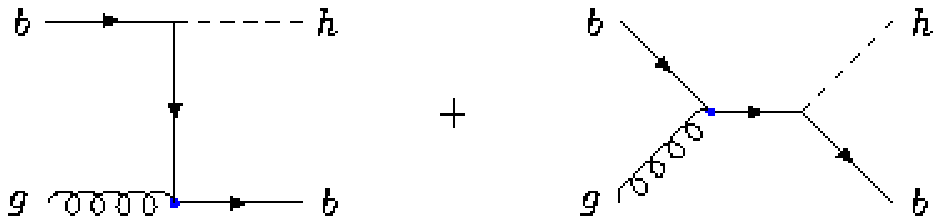}
   \caption{\it Associated production of the Higgs boson and a single hight-$p_T$ bottom quark.\label{feybis}}
 \end{minipage}
\end{figure}

The Feyman diagram of the production of a Z boson and a heavy-quark
jet via $gQ\rightarrow ZQ$, shown in figure \ref{fey},   
can probe the parton distribution function (PDF)
of the b quark, because the initial state involves a b quark from the
sea. In addition, this channel is very important because it is a
possible background for $gb\rightarrow hb$, where the Z boson and the Higgs boson decay to the same final
state ($b\bar b, \tau^+\tau^-, \mu^+\mu^-$), see figure \ref{feybis} \cite{maltoni2}.

\begin{figure}[!h]
\begin{center}
\epsfig{figure=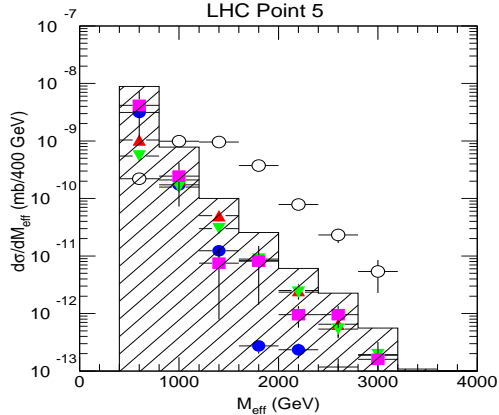,height=5.5cm,width=6.5cm} 
\end{center}
\caption{\it $M_{eff}$ distribution for the signal (open circles) and
  for the sum of all Standard Model backgrounds (histogram); the
  latter includes $t\bar t$ (solid circles), W+jets (triangles),
  Z+jets (downward triangles), and QCD jets (squares), \cite{tdr}.  
\label{fig:susy}    }
\end{figure}

The Z boson with a heavy-quark jet is a possible background
for SUSY events characterized by multiple jets, leptons, and missing
transverse energy $E_{T}^{miss}$. Looking at the plot of the
``effective mass'' shown in figure \ref{fig:susy}, defined as the scalar sum of the missing energy and
the transverse momenta of the hardest jets, the Z+jets channels
represents one of most significant Standard Model background.

Z+jet events will be used to calibrate the calorimetric jet energy
measurements, profiting from the high statistics and a relative low and
well known background.

Performing an ``in situ calibration'', it will be possible to
calibrate the calorimeters using jets reconstructed in the experiment.
This aspect will not be included in this note, for more details see
\cite{santoni} and \cite{gupta}. 

\section{Study of the Parton Distribution Function (PDF)}
At the LHC, every measurement will be deeply affected by the
uncertainties related to the knowledge of PDFs, since the production 
cross section in a proton-proton collision is given by 
the convolution of the partonic cross sections with the PDFs.


Since the kinematic space accessible at the LHC will be much
broader than at previous experiments, it will possible to study 
the PDF's both at the Electroweak mass scale (W
and Z Mass), where the low x-gluon contribution will be more
significant, and at the TeV scale where the high x-gluon will be the
dominant contribution.

\begin{figure}[!h]
\begin{center}
\epsfig{figure=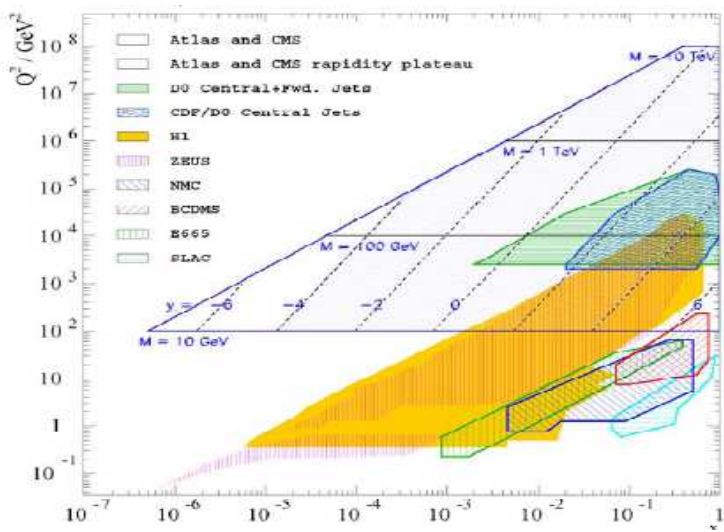,height=7cm,width=10cm} 
\end{center}
\caption{\it LHC kinematical plane in comparison with
  other experiments.
\label{plane}}

\end{figure}

Figure \ref{plane} shows the kinematic region in the
plane ($x,Q^2$) at different values of $\eta$ and mass scale for
different experiments. It is important to underline that the plane
for LHC will be greater than at previous experiments and 
that there will be some regions accessible both by LHC and the
experiments at HERA representing a good
cross check for the first measurements at LHC.

In this paper we will focus on the possibility to study the b-PDF
using the Z boson production in association with a jet.

\begin{figure}[h!]
\begin{center}
\epsfig{figure=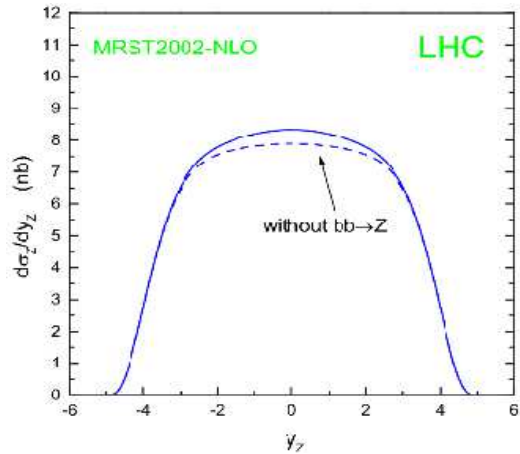,height=6cm,width=7cm} 
\caption{\it Contribution of the $b\bar b \rightarrow Z$ to the
  total Z production at LHC as function of the rapidity,
  see \cite{workshop}. 
\label{pdf2}}
\end{center} \end{figure}

The knowledge of the b PDF is directly related to the uncertainties on
the Z production cross section. At LHC, the contribution to the
total Z production from $b\bar b \rightarrow Z$ will be about 5\%, see
plot in figure \ref{pdf2}, consequently a precision of $\approx$ 1\% in the measurement of the Z
boson cross section will imply a precision of the b PDF of at least
$\approx$ 20\%.
The results obtained by HERA
are now far from this level of accuracy, see \cite{workshop}.

\begin{figure}[h!]
\begin{center}
\begin{tabular}{c c}
\epsfig{figure=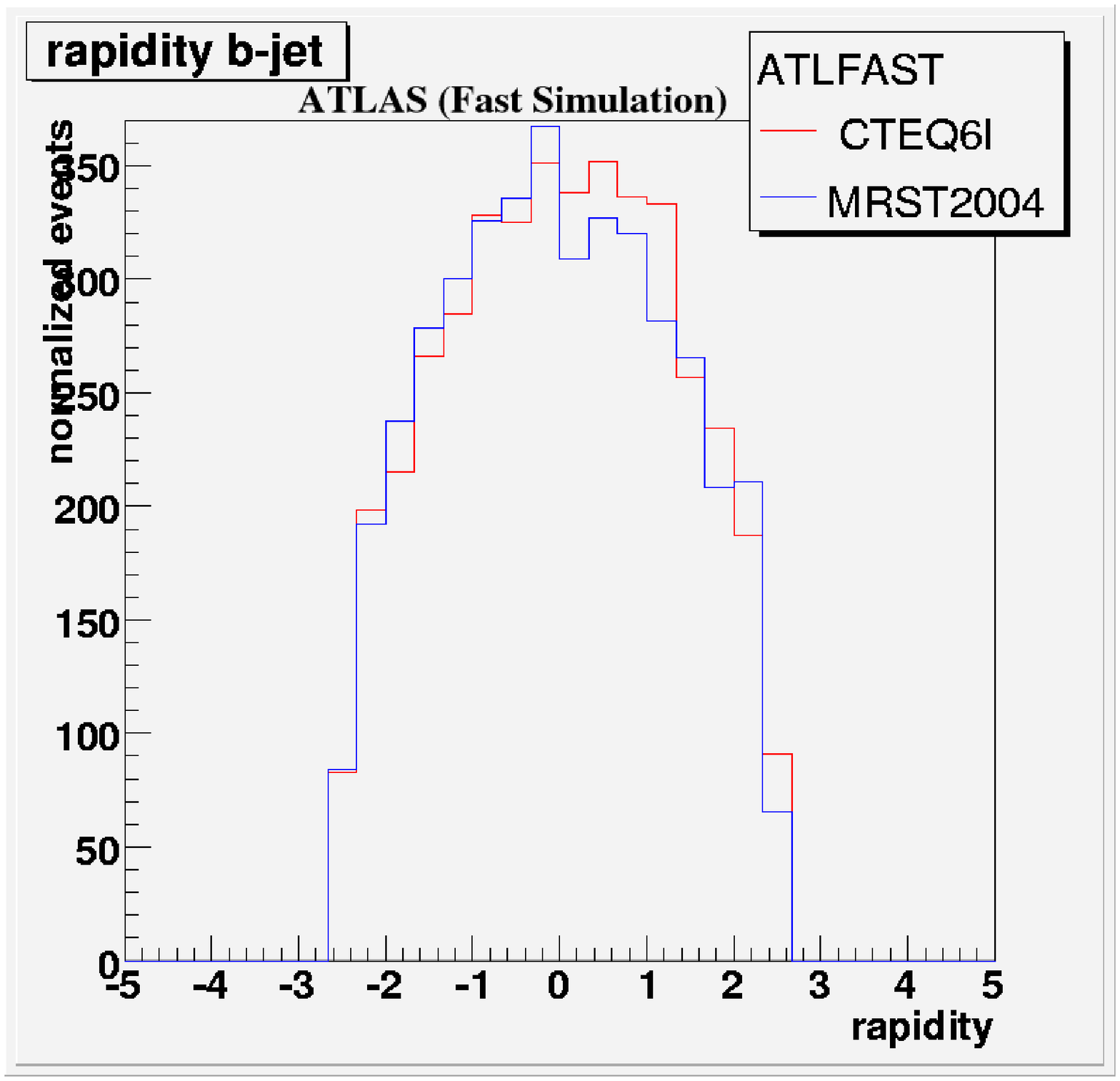,height=6cm,width=7cm} &
\epsfig{figure=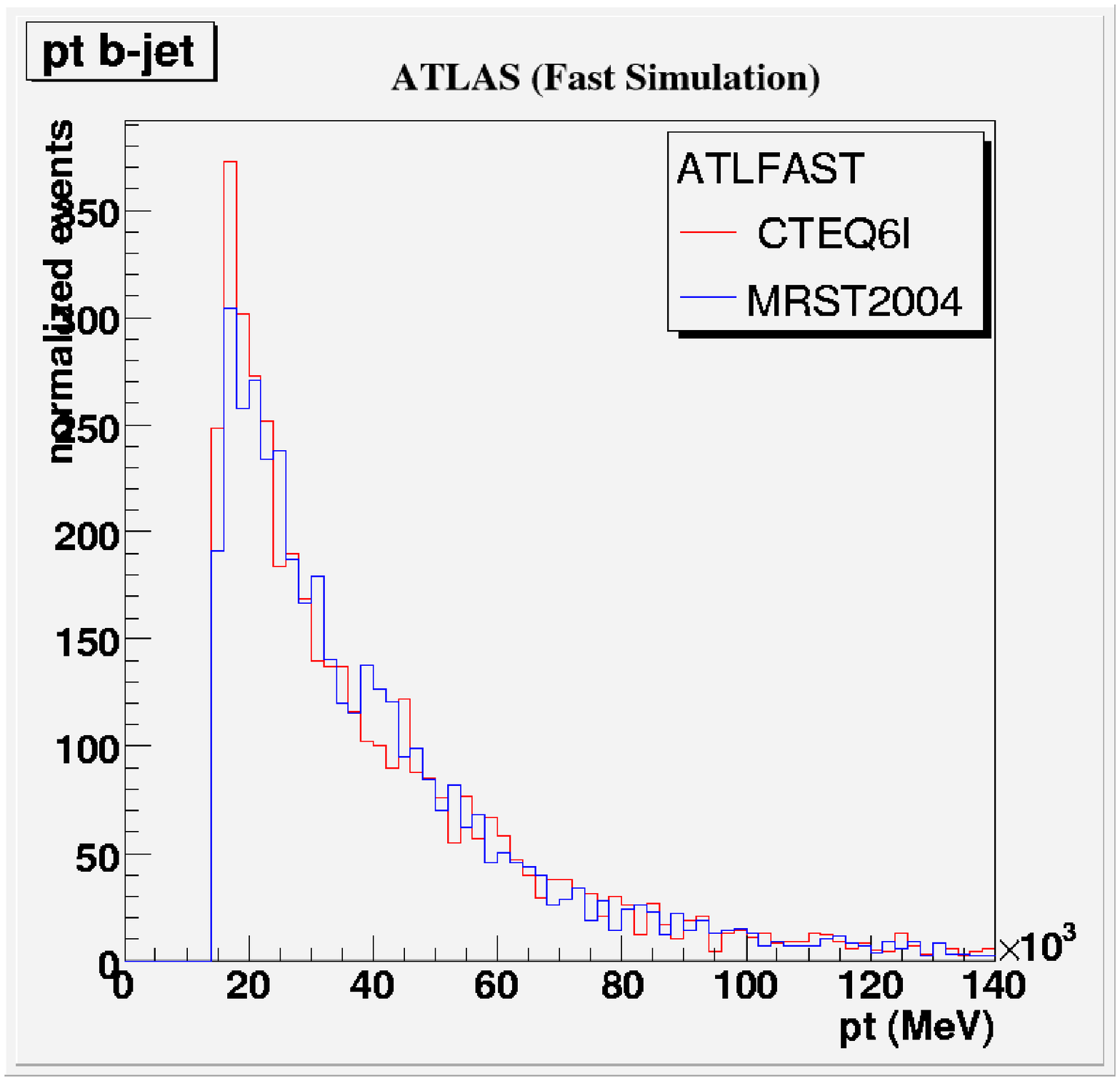,height=6cm,width=7cm} \\
(a) & (b)\\
\end{tabular}
\caption{\it (a) $\eta$ distribution of the b-jets in Zb events (ATLFAST) 
(b) $p_T$ distribution of the b-jets in Zb events (ATLFAST). 
\label{pdf1}}
\end{center} \end{figure}

  
The sensitivity to the b PDF is enhanced in processes such 
as those shown in figure \ref{fey}, involving the
production of a Z boson in association with a b quark.
At LHC we expect to have a very large statistics of such events, 
providing a measurement that
will mainly depend on systematic uncertainties.
Some preliminary studies on the $p_T$ distributions for the jets for
different sets of events generated with different PDFs have shown that
we could have differences in the low $p_T$ region of the order of
 $\approx$ 5\%, see plots in figure \ref{pdf1} for the $\eta$ and
$p_T$ distributions. 
The measurement will be sensitive if the systematic
uncertainties can be kept below this level.

As said before, the main leading order contribution will be due to the     
$gQ\rightarrow ZQ$ channel.

\begin{figure}[h!]
\begin{center}
\begin{tabular}{c c}
\epsfig{figure=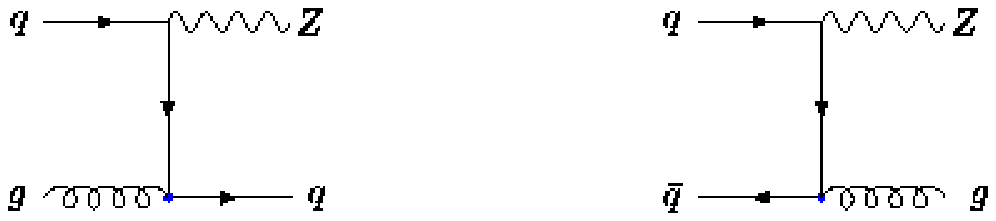,height=2cm,width=5cm} &
\epsfig{figure=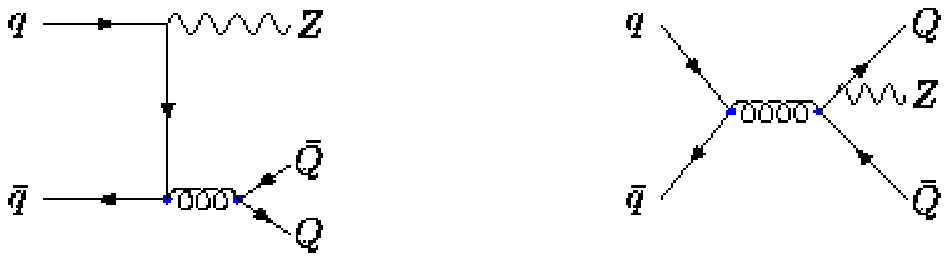,height=2cm,width=5cm} \\
(a) & (b)\\
\end{tabular}
\caption{\it (a)Feynman diagrams for Zj production via $gq\rightarrow
  Zq$ and $q\bar q\rightarrow Zg$, where q are light quarks.(b)
Feynman diagrams for $q\bar q\rightarrow ZQ\bar Q$.
\label{fey2}}
\end{center} \end{figure}

The possible sources for ZQ production, see table \ref{tab:c}, are:
\begin{itemize}  
\item{$gQ\rightarrow ZQ$\\
calculated at the leading order and at the next-to-leading order via
the subprocesses:
\begin{itemize}
\item{$gQ\rightarrow ZQ$,}
\item{$qQ\rightarrow ZqQ$,}
\item{$gQ\rightarrow ZgQ$,}
\item{$gg\rightarrow ZQ\bar Q$,}
\end{itemize}
}
\item{$q\bar q\rightarrow ZQ\bar Q$\\
calculated only at the leading order with one Q out of detector
acceptance.
Figure \ref{fey2} shows the Feynman diagrams, where one of
the Q quark is missed or two Q quarks are together in a single jet.
Nevertheless, this process is more significant at Tevatron than
at LHC.   
}
\item{$q\bar q\rightarrow Zg, gq\rightarrow Zq$\\
calculated at the leading and next-to-leading order for events
Z+j or Z+jj.
}
\item{$gg\rightarrow ZQ\bar Q$\\
this process allow the emission of one Q collinear to the beam yielding
to ZQ in the final state.
This approach, with $m_Q$ non equal to zero, however, has some problems
with the calculations, in fact the perturbation theory results less
convergent ($\alpha_Sln(M_Z/m_Q)$ instead of $\alpha_S$) and it is
much more difficult to obtain the next-to-leading order correction.
This case is not included in this study.  
}
\end{itemize}

Table \ref{tab:c} gives the cross sections of the processes described
above at the leading and next-to-leading
order, from the study \cite{maltoni1}. 
We defined the process Zb as signal and the other processes as possible backgrounds. 

The motivations to study this process at LHC appear more evident from
the comparison with the Tevatron cross sections. First of all, we notice
that the total cross section for Z+b production is about 1090 pb at LHC,
a factor of about 50 larger than at the Tevatron. We also remark that
the contribution to Z+b production from $gb\rightarrow Zb$,
which is the most important for the sensitivity to the b PDF,
is much more significant at the LHC.
Finally, at the LHC, the relative importance of processes other than Z+b
(such as Zc, Zq and Zg) is less relevant then at the Tevatron;
in addition, the probability of mis-tagging a light jet as a heavy quark
is smaller, therefore the LHC provides a cleaner environment for the
extraction of the Z+b signal.
And the total cross section for Zc production is more important
at Tevatron (about 70$\%$ greater than Zb) than at LHC (about 35$\%$ 
greater than Zb). 



\begin{table}[!h]
\begin{center}
\begin{tabular}{rrr}
\hline
{ {Cross} {Section(pb)}}
  & { {Tevatron}}
  & { {LHC}}\\
\hline
\bf Process &  &{ \bf ZQ inclusive}\\
 \hline
&& \\
$gb\rightarrow Zb$& 13.4 $\pm$ 0.9  $\pm$ 0.8 $\pm$ 0.8 &$1040^{+70+70+30}_{-60-100-50}$ \\
$q\bar q\rightarrow Zb\bar b$& 6.83 & 49.2 \\ \hline
&&\\
$gc\rightarrow Zc$& $20.3^{+1.8}_{-1.5}\pm 0.1^{+1.3}_{-1.2}$& $1390
\pm 100^{+60+40}_{-70-80}$ \\
 $q\bar q\rightarrow Zc\bar c$& 13.8& 89.7 \\ \hline
  & &{\bf Zj inclusive} \\ \hline
&&\\
$gq\rightarrow Zq$ , $q\bar q\rightarrow Zg$ &$1010^{+44+9+7}_{-40-2-12}$
 & $15870^{+900+60+300}_{-600-300-500}$ \\
\end{tabular}
\caption {Next-to-leading-order inclusive cross section (pb) for Z Boson production in association
  with heavy-quark jets at the LHC ($\sqrt s = 14 TeV pp$) and
  Tevatron ($\sqrt s = 1.96 TeV p\bar p$).
  The calculations are limited to the case of a jet in a range
  $p_T >15 GeV$ and $|\eta| <2.5$ (LHC) or $|\eta| <2.0$ (Tevatron).
The labels on the columns mean: ZQ = Z plus one jet, which contains a
  heavy quark, while Zj = Z plus one light jet, which
  does not contain a
  heavy quark, \cite{maltoni1}. The process Zb represents the signal
  of this study while all the other processes some possible backgrounds. }
\label{tab:c}
\end{center}
\end{table}

Finally, due to a relatively smaller contribution of $q\bar
q\rightarrow ZQ\bar Q$ and, for Zj events, a smaller probability of
mis-tagging a light jet as a heavy quark, LHC provides a clearer environment for
the extraction of the Zb signal.

D0 \cite{do} has provided some measurements of the cross
section ratio: (Z+bjet)/(Z+jet) within
a kinematical region similar to the one in this note.
The result is in agreement with the same ratio obtained in CDF, \cite{cdf},
and with the NLO calculations reported in \cite{maltoni1}.

In the following calculations, the values of the cross sections, 
shown in table \ref{tab:c}, will be used.

\section{Simulated Data Sample}
All the samples used have been generated using the PYTHIA MonteCarlo
package \cite{Pythia} and \cite{pythia}.
The contribution from virtual photon production (Drell-Yan) was
switched off, and only the decays of the Z boson to muons
were enabled.

The simulation of the data has been performed using, for the detector
description, the Rome Layout,
defined for the Atlas Physics Workshop held on June 2005 in Rome.

In detail we used:
\begin{itemize}
\item{$Z+jet$ with $Z\rightarrow\mu\mu $ about 500k events ($umich.004290.reco1004.ZJET010020\_mumu$)}
\item{$W+jet$ with $W\rightarrow lepton +\nu $ about 300k events\\ 
($umich.004285.reco1004.WJET010020\_lepnu$ for the
  systematics studies)
}  
\end{itemize}

Only the decays $W\rightarrow \mu\nu $ and $Z\rightarrow\mu\mu$ have
been considered. 
The W +jet process, with $W\rightarrow \mu\nu $, has been used to estimate the
systematics errors due to the mis-tagging.

The events have been analyzed using the combined Ntuples produced by
RecExCommon, the official Atlas software reconstruction package inside
ATHENA.
In this package there are dedicated algorithms for the reconstruction of different
trigger-objects like jets, leptons, photons or B-tagging algorithms to
identify jet flavor.
In particular, for the B-tagging we used the Cone algorithm, explained
in detail in section \ref{btag}.
While for the muon identification, there are two packages: MOORE and MUID, that perform the
muon reconstruction in the Muon Spectrometer, including the corrections due to the multiple
scattering and the energy loss in the inert material as described in
detail in \cite{moore}, \cite{moore2} and \cite{moore3}.

\section{Event Selection Criteria}

The selection reported in this work, defined a strategy to select, in
the first place,
events with Z and one jet, then to identify in these events a Z
boson associated with a jet from a b quark.
Two different and independent b-jet identification algorithms have
been used: ``the inclusive b-tagging'' and ``the soft muon tagging'',
described below.

The signal was defined as the events containing a Z boson decaying
into a couple of muons and a b jet with a $p_{T} \geq 15 GeV$ and $|\eta| \leq 2.5$. 
The background samples containing respectively a c-jet within the same cuts, or a jet
originating from a light quark or a gluon in the same range, were
considered separately. The NLO cross-sections computed in table \ref{tab:c}
were used for the signal and for these two classes of background,
while the cross-section given by PYTHIA was taken for the events with
a jet from light quark or gluon.

In the section below the selection performaces of Z+jet events with the two
b-tagging algorithms are presented. 

The quality of the event selection has been estimated using the
ratio variables: {\sl efficiency} and {\sl purity}.
The efficiency is defined as the ratio between the number of selected
signal events over the total number of events while the purity is given by
the ratio between the number of events with a b-quark and the number
of selected signal events.

The number of expected events is estimated using the formulas
(1), where we used the cross sections at the NLO,
taking into account the different efficiencies of selection: $\epsilon_{acc}$
due to the geometric acceptance, $\epsilon_{cuts}$ for the cuts applied
for the di-muon invariant mass and finally $\epsilon_{x}$ for the
selection of the sample with the jet of a particular flavor.

\begin{eqnarray}
N_b &=  &\sigma_b^{table}\cdot BR(Z\rightarrow\mu\mu)\cdot \mathcal L \cdot \delta t \cdot
\epsilon_{acc}\cdot \epsilon_{cuts}\cdot \epsilon_b \cr
N_c  &= &\sigma_c^{table}\cdot BR(Z\rightarrow\mu\mu)\cdot \mathcal L \cdot \delta t \cdot
 \epsilon_{acc}\cdot \epsilon_{cuts}\cdot \epsilon_c  \cr
N_{oth}  &= & \sigma_{other}^{Pythia}\cdot
BR(Z\rightarrow\mu\mu)\cdot \mathcal L \cdot \delta t \cdot
 \epsilon_{acc}\cdot \epsilon_{cuts}\cdot\epsilon_{oth} \cr
\label{stima}
\end{eqnarray}

An estimation of the number of events expected for $30fb^{-1}$ of
integrated luminosity has been provided. 

\subsection{Z+jet Events}


The selection of these events will be primarly affected by the trigger efficiency. 
The trigger efficiency (2mu20 trigger menu) is high and about 95\%
\cite{hlt}, thanks to the presence of one or more high $P_T$ muons in
the event. This is the major advantage of this channel with respect to the other
events with b quarks.

Concerning the efficiency of the muon reconstruction algorithm, MUID, has a selection
efficiency in the $p_{T}$ range of the muons from Z boson decay of more
than 95\% for $|\eta| <2.5$.


The sample of Z+jet events was selected requiring two high $p_T$ muons satisfying
the following kinematic cuts, and at least one jet: 

\begin{itemize} 
\item{two muons of opposite charge, }
\item{both muons with $p_T \geq 20$ GeV, }
\item{di-muon invariant mass in the range $70 \leq M^{\mu\mu} \leq 110$ GeV, }
\item{and one jet.}
\end{itemize}
   
In the case where there are more than two muons satisfying the cuts
all the possible combinations for
di-muon invariant mass have been taken into account, the best one has
been chosen.

To define the geometric acceptance of the detector and
the coverage of reconstruction software, these additional cuts have been
required:
\begin{itemize} 
\item{two muons in $|\eta|\leq 2.5$,   }
\item{one jet in $|\eta|\leq 2.5$ and $pT > 15 GeV$. }
\end{itemize}

\begin{figure}[h!]
\begin{center}
\begin{tabular}{c c}
\epsfig{figure=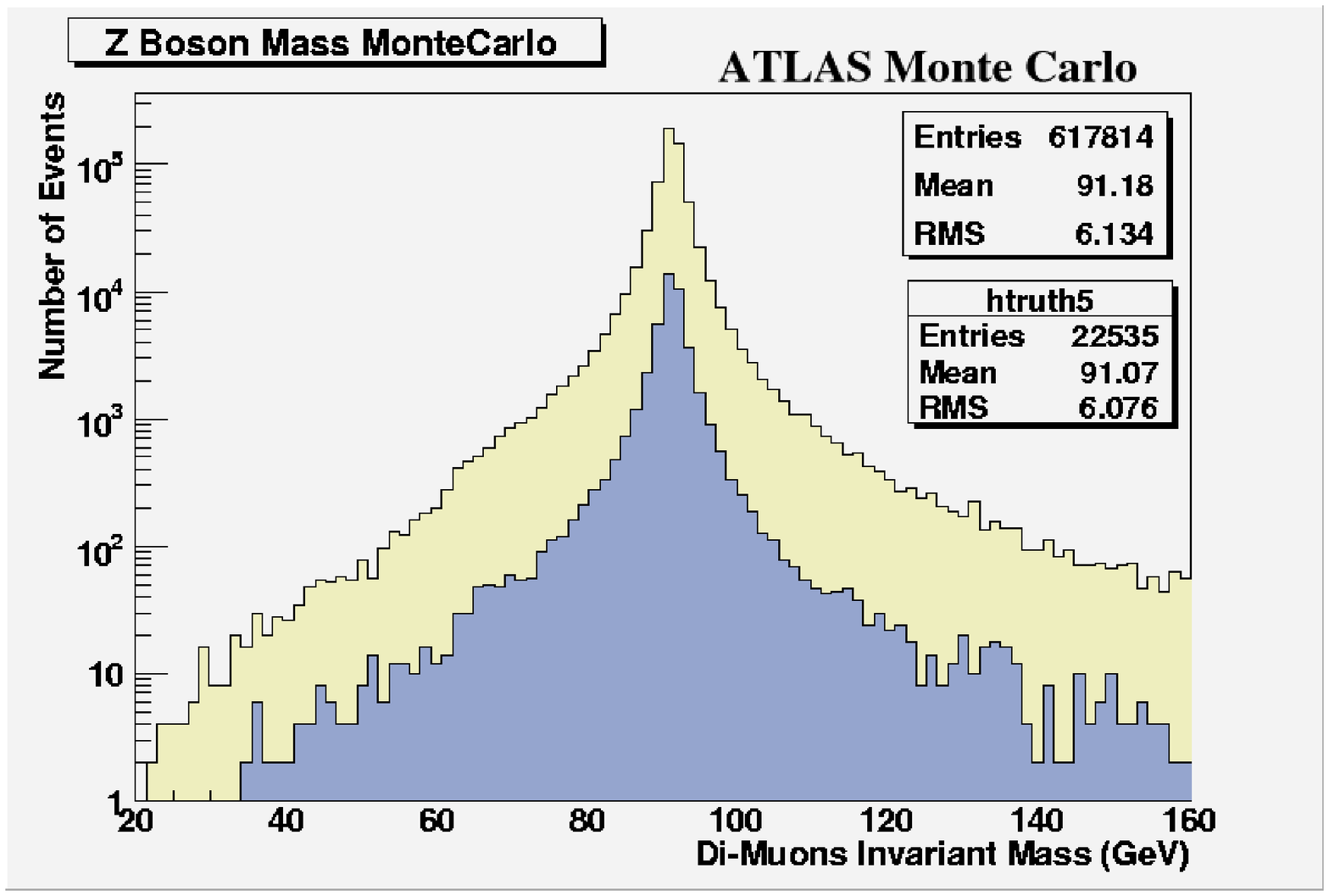,height=7cm,width=7cm} &
\epsfig{figure=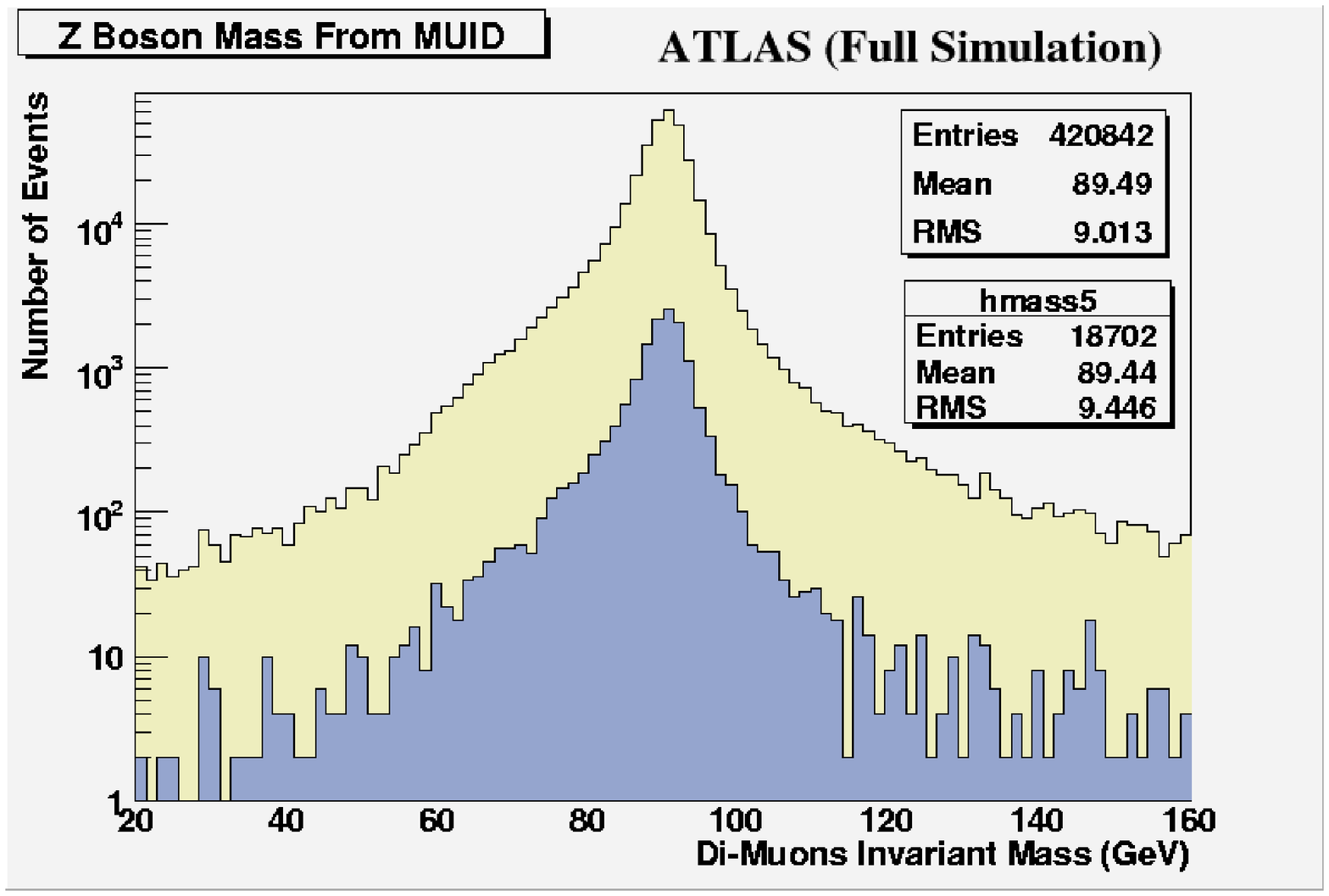,height=7cm,width=7cm} \\
(a) & (b)\\
\end{tabular}
\caption{\it (a) Di-muon invariant mass from MonteCarlo information,
  (b) same plot but obtained with the reconstruction package MUID.
The same cuts are applied in the plots, requiring two muons with
  $\eta<2.5$ and with a $P_{T} > 20$ GeV. The dark (blue)
  histogram represents the events with b-quark. 
\label{fig::mass}}
\end{center} \end{figure}

Figure \ref{fig::mass} shows the di-muon invariant mass plots
obtained by MonteCarlo Truth (a) and by MUID (b) respectively after
these cuts. 

\subsection{Z+b-jet Events \label{zjet}}

\subsubsection{Secondary Vertex b-Tagging \label{btag}}
The inclusive b-tagging algorithm is used to identify
the jet of b flavor, see \cite{b} for additional information about
the algorithm.


The jet is identified with the Cone algorithm with $\Delta R =
\sqrt{(\Delta \phi)^2 + (\Delta \eta)^2} = 0.7$, and requiring:
\begin{itemize}   
\item{at least one track selected in the tracker for jet tagging,}
\item{cut on the b-tagging weight $w_t >3$. Tracks from B-hadrons decays are
  expected to have on average a large and positive transverse impact
  parameter $a_0$ and a large and positive longitudinal impact
  parameter $z_0$, but less discriminating. Combining the
  longitudinal and trasverse significance in the function: $w_t=
  P_b(S_{a_0,z_0})/P_u(S_{a_0,z_0})$ and applying the previous cut on
  the likehood function is possible to discriminate b-jet from light jets. }
\end{itemize}

\begin{figure}[h!]
\begin{center}
\begin{tabular}{c c}
\epsfig{figure=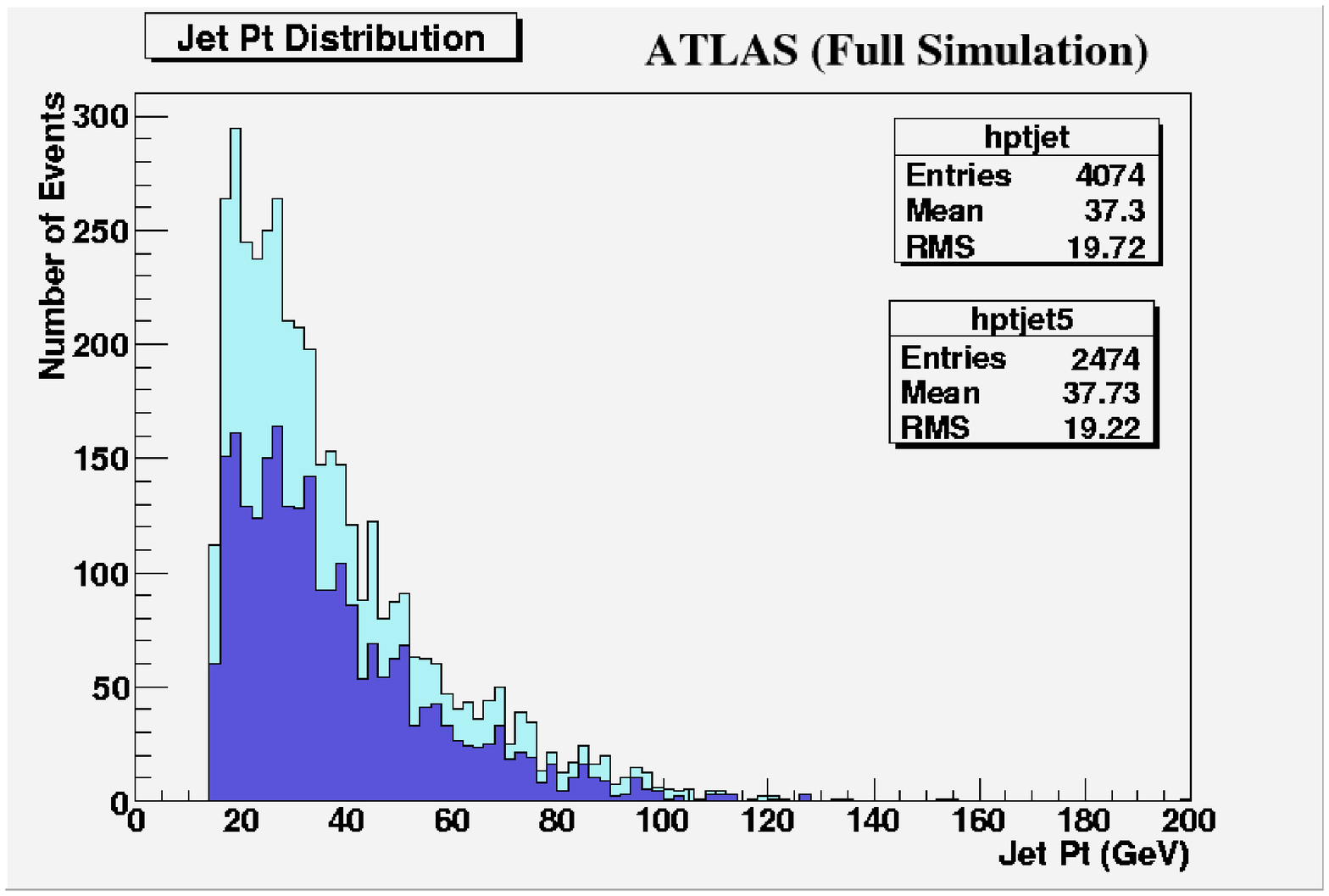,height=7cm,width=7cm} &
\epsfig{figure=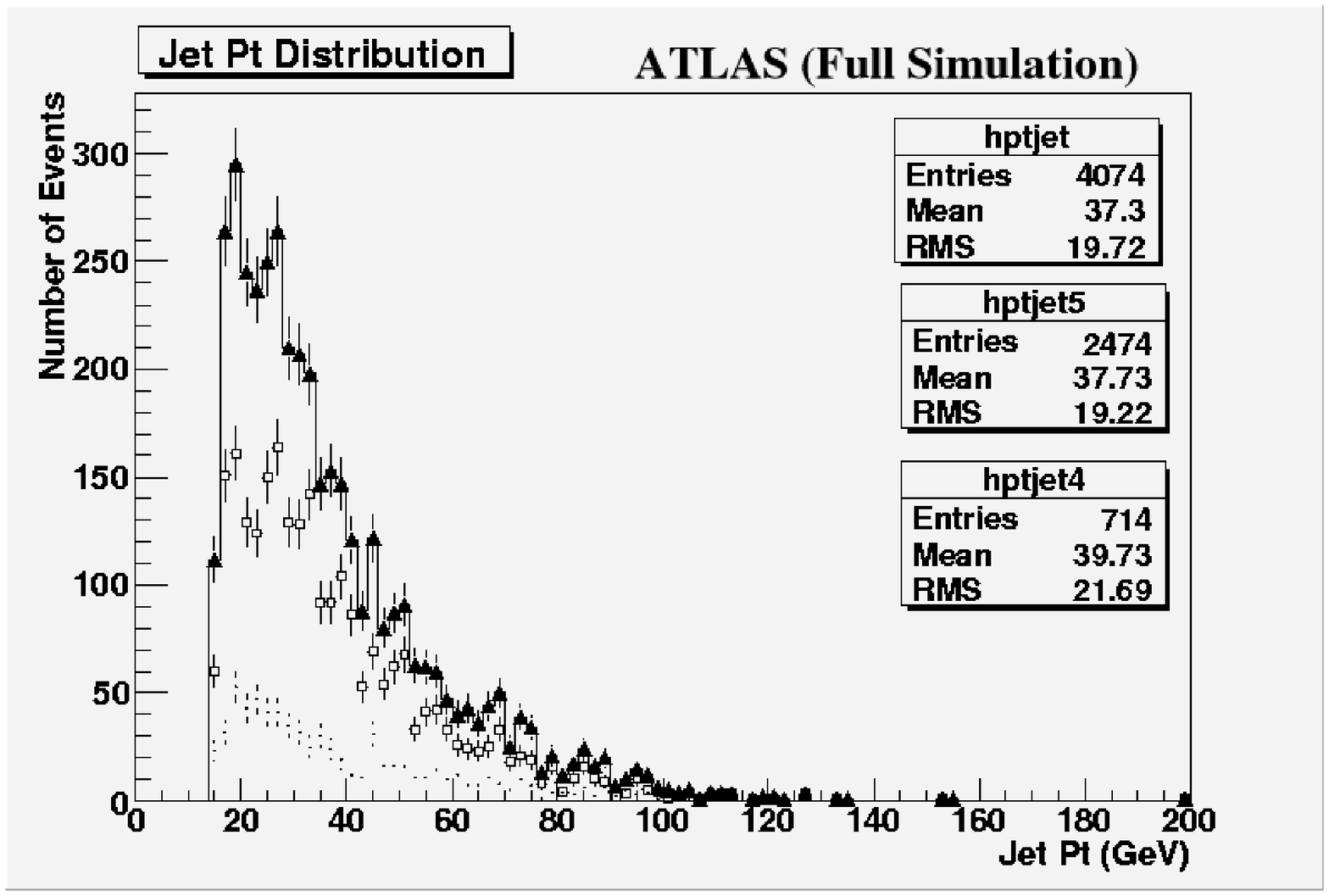,height=7cm,width=7cm} \\
(a) & (b)\\
\end{tabular}
\caption{\it (a) $P_{T}$ distribution of jets obtained after the cuts on the
  di-muon invariant mass and the b-tagging jet selection. The dark area represents the tagged b-jets. 
  (b) $P_{T}$ spectrum of jets obtained after the cuts on the
  di-muon invariant mass and the jet. The b (squares - hptjet5) and c
  (dotted line - hptjet4) jets are shown
  separately with respect the total jets sample (line and black
  triangle - hptjet).
\label{btag}}
\end{center} \end{figure}

\begin{figure}[h!]
\begin{center}
\begin{tabular}{c c}
\epsfig{figure=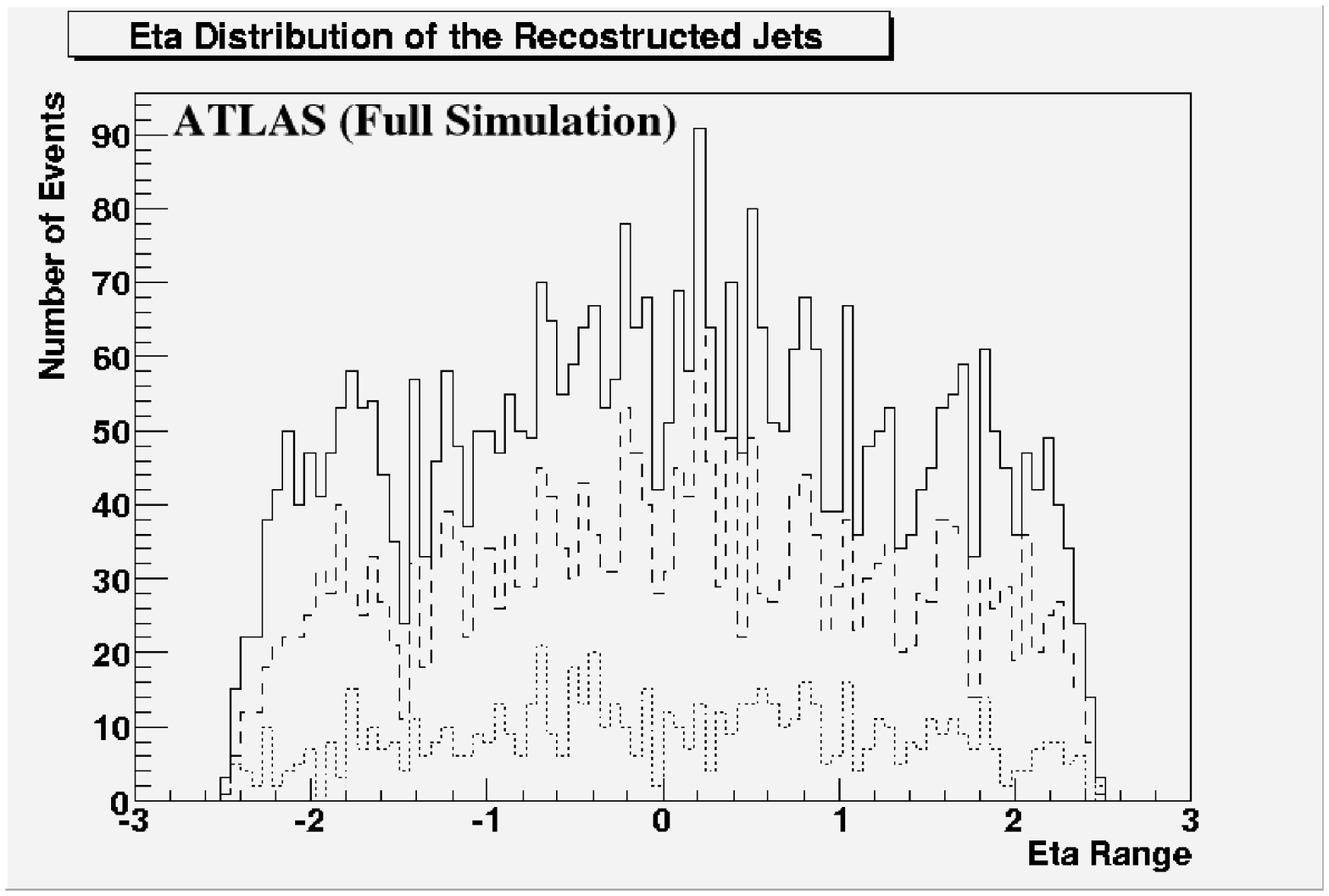,height=7cm,width=7cm} &
\epsfig{figure=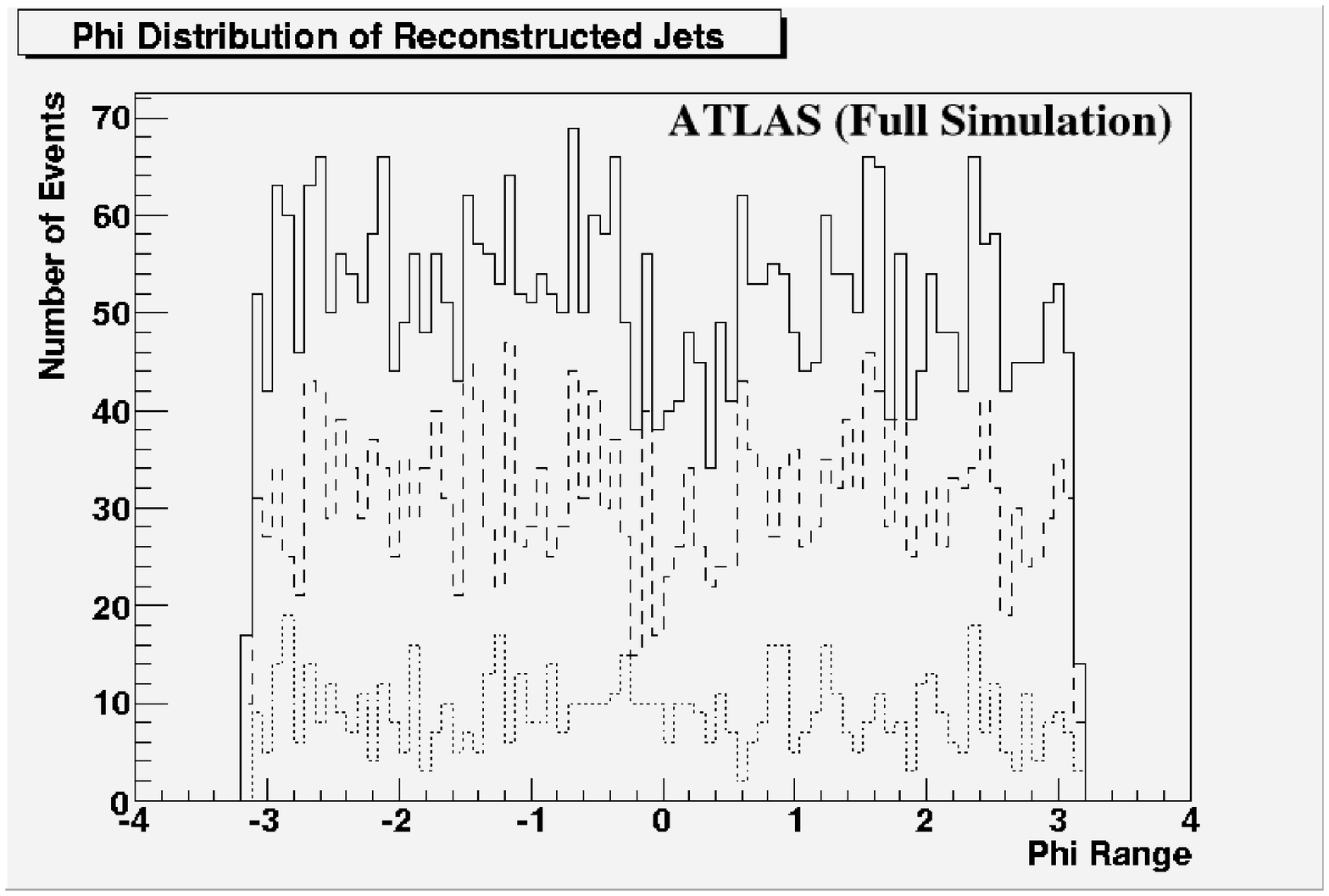,height=7cm,width=7cm} \\
(a) & (b)\\
\end{tabular}
\caption{\it (a) Eta distribution of jets obtained after the cuts on the
  di-muon invariant mass and the b-tagging jet selection. The b (dash line) and c (dotted line) jets are shown
  separately with respect the total jets sample (full line). 
  (b) Phi distribution of jets obtained after the cuts on the
  di-muon invariant mass and the b-tagging jet selection, shown with
  the same labels. 
\label{btag_phi}}
\end{center} \end{figure}

The obtained purity and the expected number of events are reported in table \ref{tab:b}.
Figures \ref{btag} show the $P_{T}$ spectrum of jets in the selected
events, while the figures \ref{btag_phi} the distributions of
the eta and phi variables for the jets selected by the b-tagging
algorithm. 
The fact that b-tagged jets tend to have a higher $P_{T}$ implies that
the effiency and purity of the selected sample improve after the
additional cut of the ${p_T>15GeV}$. 
The relative quantity of b-jets with
respect of the total number of selected jets is about
60\% while the contamination from c-jets is about 20\% in all the
$P_{T}$ range, as shown in detail in the table \ref{tab:bconta} .

\begin{table}
\begin{center}
\begin{tabular}{cc}
\hline
\bf{B-TAGGING} & \\ \hline
  \bf{CUTS}
  & \bf{Efficiency}\\

\hline
\cline{1-2}
$\mu$ in $|\eta|<2.5$ & 49.9$\%$\\ \hline
$p_T>20 GeV$ and $70<Mass<110 GeV$ & 59.5$\%$\\ \hline
B-tagging Algo & 56.2$\%$\\ \hline
\cline{1-2}
 \bf{Purity} & 60.7$\%$\\ \hline
\bf{Number of b Events}& 176642\\\hline
\bf{Number of Background Events}& 204265\\\hline
\end{tabular}
\caption{ The efficiencies, the purity and the number of estimated events with the full 
simulation for the b-tagging algorithm.
The number of events is calculated for 30 $fb^{-1}$ of integrated luminosity. The b,c jets selected
satisfied the cuts on $p_{T}$ and $\eta$: ${p_T>15GeV}$ and $|\eta|\le 2.5$ } 
\label{tab:b}
\end{center}
\end{table}

\begin{table}
\begin{center}
\begin{tabular}{|ccc|}
\hline
\bf{B-TAGGING} & & \\ \hline
  \bf{$P_{T}$ Range in GeV}
  & \bf{ b jets \% }  & \bf{ c jets \% }\\
\hline
\cline{1-3}
15-25 & 54.2 & 17.2\\ \hline
25-35 & 63.2 & 15.4\\ \hline
35-45 & 75.6 & 14.6\\ \hline
45-55 & 58.2 & 20.3\\ \hline
55-65 & 58.7 & 21.4\\ \hline
65-75 & 59.8 & 21.1\\ \hline
75-200 & 47.8 & 24.7\\ \hline

\end{tabular}
\caption{ Population of b-jets and c-jets, divided in $P_{T}$ range, after the
  B-Tagging selection and the additional cuts on the jets of $\eta$: ${p_T>15GeV}$ and $|\eta|\le 2.5$} 
\label{tab:bconta}
\end{center}
\end{table}

A first estimation of the ratio between the Z+bjet cross
section over the total production cross section of Z+jet is given,
taking into account the results obtained for the 30 $fb^{-1}$ of
integrated luminosity and only for the inclusive B-Tagging selection.

The cross section ratio is defined as the ratio of the number of events
obtained after the b-tagging selection over the total number of Z+jet
events multiplied for the ratio of the selection efficiencies and the
ratio of the purities of the samples.
In detail:

\begin{equation}
{Ratio} = {\frac{\sigma_{Z+bjet}}{\sigma_{Z+jet}}} =
  {\frac{N_b}{N_{tot}}}\cdot {\frac{\epsilon_{1jet}}{\epsilon_{1bjet}}}
      \cdot  {\frac{f_{B\rightarrow b}}{1}}
  = {\frac {4074}{114881}} \cdot {\frac{0.6}{0.6}} = 0.035
\end{equation}

where $\epsilon_{onejet}/\epsilon_{onebjet} = 1/\epsilon_{BTag} \approx 0.6$,
see \cite{b}, ${f_{B\rightarrow b}} = 1- f_{C\rightarrow b} -
f_{lightquark\rightarrow b}\approx 0.6$.
The statistical error $\sigma (stat)$ is less than $10^{-4}$ due to the high
statistics and the systematic error $\sigma (syst)$ is still under
evaluation, it is expected to be $\leq 10\%$. 
The value of the ratio obtained by the D0 measurement for
$180 pb^{-1}$, is: $R = 0.019 \pm 0.005 (stat)$
\cite{do}.

\newpage

\subsubsection{Soft Muon Tagging}
The soft muon tagging provides an independent way from the usual
b-tagging algorithm to identify the b-jet using the muon from the
semileptonic decay of a b-meson. 

At this stage, we have only looked for a third muon in the Muon
Spectrometer that has been reconstructed by MUID.
Typically, the muon inside the b-jet can reach the middle station of the Muon Spectrometer
to be reconstructed.
The efficiency of this selection is intrinsically low, depending on the
few muons that can reach the Spectrometer after have lost 2-3 GeV in
the calorimeters, but the purity obtained is rather good, introducing
a cut on the transverse momentum of the muon the purity increases.
The simple request of a third muon in the soft muon-tagging implies
an underestimation of the relative background.

In the table \ref{tab::mu} the purity and the number of expected
events for 30 $fb^{-1}$ of integrated luminosity are given.

\begin{figure}[!h]
\begin{center}
\epsfig{figure=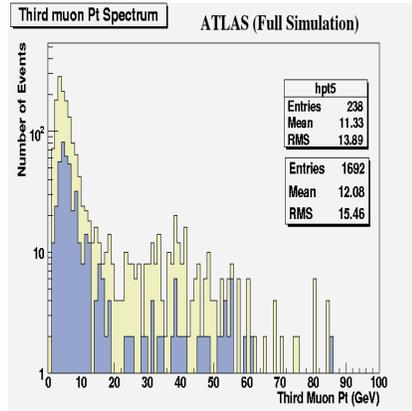,height=5.5cm,width=5.5cm} 
\end{center}
\caption{\it Third muon $P_{T}$ distribution obtained after all cuts and 
requiring at least one jet.
The muon $P_{T}$ is reconstructed in the Muon Spectrometer using the
  MUID reconstruction algorithm. 
\label{fig:softmu}    }
\end{figure}

\begin{table}
\begin{center}
\begin{tabular}{cc}
\hline
\bf{SOFT MUON TAGGING} & \\ \hline
  \bf{CUTS}
  & \bf{Efficiency}\\\hline
\cline{1-2}
$\mu$ in $|\eta|<2.5$ & 49.9$\%$\\ \hline
$p_T>20 GeV$ and $70<Mass<110 GeV$ & 59.5$\%$\\ \hline
Third Muon Algo & 7.2$\%$\\ \hline
\cline{1-2}
 \bf{Purity} & 37.2$\%$\\ \hline
\bf{Number of b Events}& 22630\\\hline
\bf{Number of Background Events}& 68088\\\hline

\end{tabular}
\caption{The Efficiencies, the purity and the number of events
  estimated with the full simulation for the soft muon
  tagging algorithm. 
 The b,c jets selected satisfied the cuts on $p_{T}$ and $\eta$: ${p_T>15GeV}$ and
  $|\eta|\le 2.5$.
The number of expected events are calculated for an integrated
  luminosity of 30 $fb^{-1}$.  
\label{tab::mu}} 

\end{center}
\end{table}

\subsubsection{Further Studies on the b Jets Discriminants}

The major source of background in our selected events is represented
by c-jets tagged as b-jets.

We have studied which variables can be used to better discriminate the signal events
with respect the background events after the b-tagging algorithms
(inclusive b-tagging and soft muon tagging) have been applied.

Taking into account the results of ZEUS, H1
and CDF, we have investigated, as possible discriminating variable, the
muon transverse momentum relative to the axis of associated jet.

The relative momentum is given by:

\begin{equation}
{p_{T}^{Rel}} = {|\vec{p_{\mu}}| \sin(\arccos(\frac{\vec{p_{\mu}}\cdot
 \vec{p_{jet}}}{|\vec{p_{\mu}}||\vec{p_{jet}}|}))}
\label{eqrel}
\end{equation}

\begin{figure}[h!]
\begin{center}
\begin{tabular}{c c}
\epsfig{figure=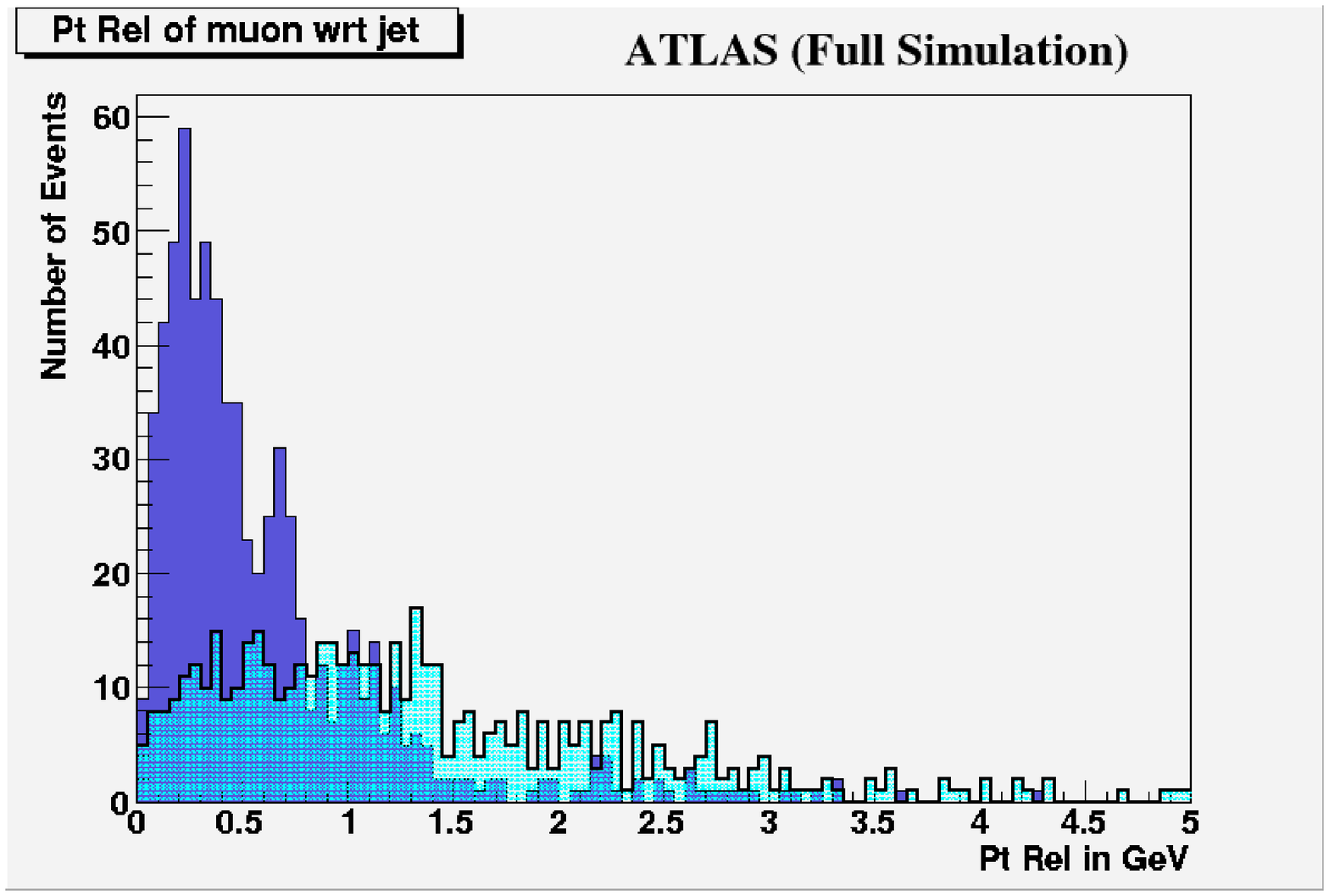,height=7cm,width=7cm} &
\epsfig{figure=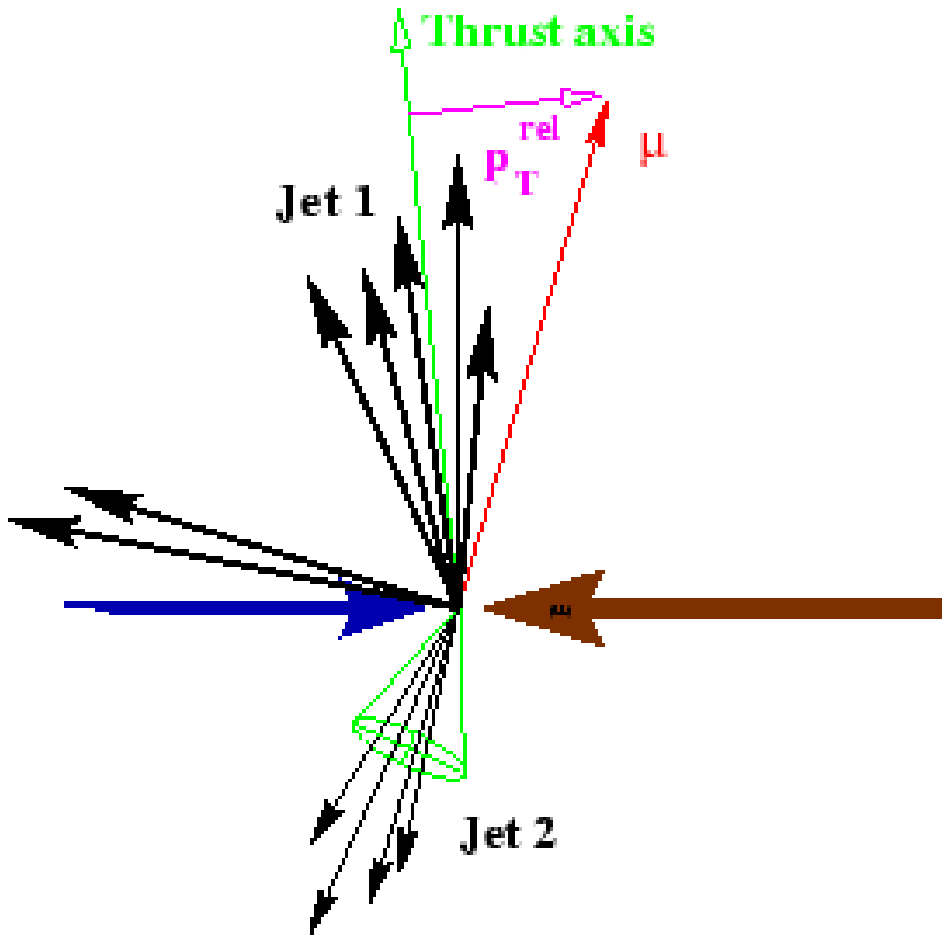,height=7cm,width=7cm} \\
(a) & (b)\\
\end{tabular}
\caption{\it (a) The $p_{T}^{Rel}$ distributions for charm (dark plot,
  less
  spread, with the maximum less than 1 GeV) and beauty (more spread
  and flat) events. This plots have been obtained using the MonteCarlo
  information for the jets and the muons.
  (b) Draft of the $p_{T}^{Rel}$. 
\label{fig:plotrel}}
\end{center} \end{figure}

The b events show to have a $p_{T}^{Rel}$ values greater than the
c events, as it is shown in the plots in figure \ref{fig:plotrel} due to
the bigger mass of the b quark.
Further variables will be studied to have a good
discriminator between signal and background for the selected events.

\section{Systematic effects on the selection}

Given the large statistics of the available data samples, the measurement 
will be limited by systematic effects which will be mainly due to the
knowledge of the selection efficiency and background estimation.

We have investigated the possibility to 
control these effects directly from data sample itself.\\
The b-tagging efficiency can be checked using $b$-enriched
samples. Based on previous experience at Tevatron
and LEP, we can expect a relative uncertainty of about 5\%.\\
The background in the selected sample is mainly due to mis-tagged jets from $c$
and light quarks.
This can be controlled by looking at the number of
tagged jets in data samples that in principle should not
contain $b$-jets at first order: $W$+jets, for example, 
are such a kind of events.

The $W$+jet events will be available 
with large statistics, moreover 
jets produced in association with $W$ boson will cover 
the full $p_T$ range of the signal.

On this bases we decided to use a $W$+jets sample (generated using PYTHIA ~\cite{Pythia} and processed with a complete 
simulation of ATLAS detector, based on GEANT4 ~\cite{geant4}) to estimate systematic 
uncertainties on $b$-tagging.\\
We analyzed only events containing a muon from $W$ decay within the following  
kinematic cuts (that are the same cuts used to select muons from $Z$ decay in the 
$Z$+jet sample):
\begin{itemize}
\item at least one muon in the event with:
\begin{itemize}
\item[-] $p_T>$ 20 GeV;
\item[-] $|\eta|<$ 2.5;
\end{itemize}
\item at least one reconstructed jet. 
\end{itemize}

\begin{figure}[!h]
\begin{center}
\epsfig{figure=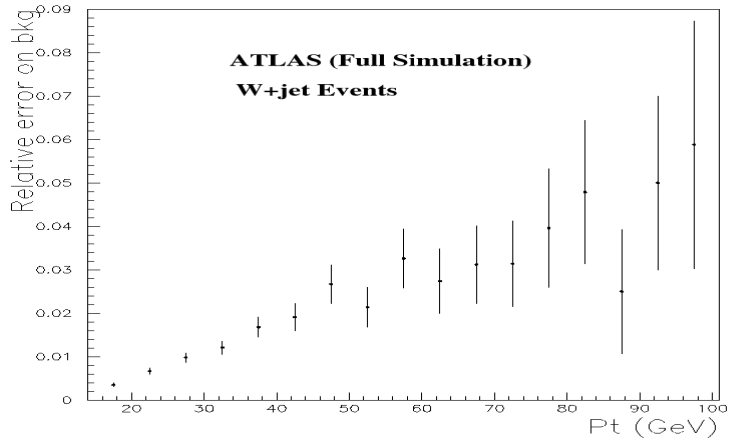,height=6.5cm,width=10.5cm} 
\end{center}
\caption{\it Systematics due to mis-tagging of $b$-jets evaluated from 
fraction of b-tagging jets in the W+jets sample: relative error on background level per 5 GeV jet $p_T$ bin.
\label{fig:err}    }
\end{figure}

After this selection on muons, we looked for events where at least one jet 
is tagged as a $b$-jet. In this way it has been
possible to calculate the number of mis-tagged jet with respect to the total 
number of jets as function of the jet $p_T$.\\
As we can see in figure \ref{fig:err} we expect to estimate the background from
mis-tagging with a relative uncertainty at the level of few percent
over the full $p_T$ range of the signal.\\
This result is in good agreement with previous results obtained 
using fast simulation on a $W$+jet sample ~\cite{proceeding}.

The systematics due to the calibrations of the detectors, according to
the same studies \cite{santoni,gupta} will be of the order of few percent, this value is
under investigation and will be checked using the data simulated for the
Computing System Commissioning CSC-06.
Further studies are in progress to define the energy scale and
resolution of the reconstructed jets.

\section{Conclusions and outlook}

In the LHC physics program, the Z+jet analysis has an important role,
it represents a clear and high statistics channel to perform studies on
the PDFs in particular on the b-PDF, and consequently to improve the
precision of the other production cross sections of topics signatures
as Higgs or SUSY.
The high statistics expected after all the analysis chain will provide
a sample of a relevant purity with a small statistical error.
The measurement will be affected principally by the systematic
uncertainties.  

Further studies will follow on the analysis of the data samples
simulated with the underlying and minimum bias events at low luminosity.

A more complete analysis concerning systematic effects and the impact of
the varies sets of pdf is already being done with the CSC data.

\clearpage



\begin{thebibliography}{40}

\bibitem{nota} 
S.Diglio, A.Tonazzo, M.Verducci, \emph{A preliminary study of Z+b
  production at LHC}, ATL-COM-PHYS-2004-078, 2004

\bibitem{maltoni1} 
J.Campbell, R.K.Ellis, F.Maltoni and S.Willenbrock, \emph{Associated
  production of a Z Boson and a single heavy quark jet},
Phys.Rev.D69:074021, 2004
\bibitem{maltoni2} J.Campbell, R.K.Ellis, F.Maltoni, S.Willenbrock,
 \emph{Higgs-Boson porduction in association with a single bottom quark}
 Phys.Rev.D67:095002, 2003

\bibitem {tdr} ATLAS Collaboration \emph{ATLAS Detector and Physics
  Performance Technical Design Report}, CERN/LHCC/1999-15, (1999)

\bibitem{santoni} R.Lefevre, C.Santoni \emph{In situ determination of
  the scale and resolution of the jet energy measurements using
  $Z^0+jets$ events}, ATLAS Internal Note, ATL-PHYS-2002-026, 2002

\bibitem{gupta} J. Proudfoot, et al. \emph{Jet Energy Scale
  Calibration in Vector Boson Events using Rome Simulation Data}, ATLAS Internal Note, ATL-COM-PHYS-2005-067, 2005

\bibitem{workshop} S.Alekhin, et al. \emph{HERA and LHC, a Workshop on the implications
  of HERA for LHC physics}, CERN-2005-014, DESY-PROC-2005-001


\bibitem{do}
DO Note 4388, The DO Collaboration,
Phys.\ Rev.\ Lett. {\bf 94} 161801 (2005);

\bibitem{cdf} CDF Collaboration, \emph{Measurement of the b Jet Cross
  Section in Events with a Z Boson in $p\bar p$ Collisions at $\sqrt s
  = 1.96 TeV$}, hep-ex/0605099, May 2006



\bibitem{Pythia} T. Sj\"ostrand, P. Eden, C. Friberg, L. L\"onnblad,
G.Miu, S.Mrenna, E.Norrbin, \emph{PYTHIA 6.2 Physics and
Manual}, Computer Physics Commun {135}, (2001) 238

\bibitem{pythia}
Atlas Collaboration Manual, \newline 
{$http://wwwinfo.cern.ch/asdoc/psdir/pythiarus/pythia$\_$rus.ps.gz$}



\bibitem{moore}
D.Adams, et al., Atlas Internal Note, ATL-SOFT-2003-008, ATL-DAQ-2003-02, 2003
\bibitem{moore2}
D.Adams, et al., ATLAS Internal Note, ATL-SOFT-2003-007, 2003

\bibitem{moore3} D.Adams, et al., ATLAS Internal Note, ATL-CONF-2003-011, 2003



\bibitem {b} S. Correard et al. \emph{b-tagging with DC1 data}, ATL-PHYS-2004-006, (2004)



\bibitem {hlt} ATLAS Collaboration \emph{ATLAS High-Level Triggers,
  DAQ and DCS Technical Proposal}, CERN/LHCC/2000-17, (2000)


\bibitem{proceeding} M. Cacciari, E. Laenen, S. Diglio, A. Tonazzo, M. Verducci et. al.
\emph{Theoretical review of various approaches in heavy quark production}, 
{CERN-2005-014} (14 December 2005)[arXiv:hep-ph/0601012-3]

\bibitem{Tricoli} A. Cooper-Sarkar, M. Dittmar, D.C. Gwenlan, H. Stenzel, A. Tricoli
\emph{LHC final states and their potential experimental and theoretical
accuracies}, {CERN-2005-014} (14 December 2005) 



\bibitem{geant4} S. Agostinelli et al., \emph{Geant4 - A Simulation Toolkit}, 
Nuclear Instruments and Methods {A 506} (2003) 250-303




\end{thebibliography}
\end{document}